# A sneak into the Devil's Colony- Fake Profiles in Online Social Networks

Mudasir Ahmad Wani[*,a]
mudasirwanijmi@gmail.com
[a] Department of Computer
Jamia Millia Islamia, New Delhi

Suraiya Jabin[a]
sjabin@jmi.ac.in

***Abstract –*** Online Social Networks (OSNs) play an important role for internet users to carry out their daily activities like content sharing, news reading, posting messages, product reviews and discussing events etc. At the same time, various kinds of spammers are also equally attracted towards these OSNs. These cyber criminals including sexual predators, online fraudsters, advertising campaigners, catfishes, and social bots etc. exploit the network of trust by various means especially by creating fake profiles to spread their content and carry out scams. All these malicious identities are very harmful for both the users as well as the service providers. From the OSN service providers' point of view, fake profiles affect the overall reputation of the network in addition to the loss of bandwidth. To spot out these malicious users, huge manpower effort and more sophisticated automated methods are needed. In this paper, various types of OSN threat generators like compromised profiles, cloned profiles and online bots (spam-bots, social-bots, like-bots and influential-bots) have been classified. An attempt is made to present several categories of features that have been used to train classifiers in order to identify a fake profile. Different data crawling approaches along with some existing data sources for fake profile detection have been identified. A refresher on existing cyber laws to curb social media based cybercrimes with their limitations is also presented.

1. INTRODUCTION

Online Social Network Analysis (OSNA) is considered as one of the most emerging research fields. An online social network (OSN) is the grouping of nodes (individuals, actors, organizations, nations, states or WebPages etc.) around the world connected by a set of links (relationships, interactions, distances, hyperlinks etc). Since its inception, OSN has changed the way people think, express, and socialize with outside world. For example to buy a new product, people find it better to look for Google reviews rather than taking a friend's advice. Currently there are umpteen Social networking sites like Facebook, Twitter, Google+, Flicker, LinkedIn, Hello etc. and almost every individual is member of one of these OSNs. These OSNs are growing rapidly in terms of the number of users and the number of connections across the different geographies. Although, OSNs are gaining universal popularity but it brings number of security and privacy challenges like spam, scam, phishing, clickjacking harassing or stalking an individual or a group, defamation, identity theft, third party personal information disclosure etc. Since user's personal, professional, social and political data is cluttered at a single place which equally attracts social spammers (cyber criminals) towards these OSNs which can be very harmful for both users as well as service providers. These cybercriminals use identity theft attacks, creating fake profiles or launching automated crawling against a number of popular social networking sites. Other reasons for creating fake profiles include advertising and campaigning, defaming a person, social engineering, fun and entertainment, data collection for research/ specialized marketing, fake traffic for blogs or websites etc.

OSNs provide huge amount of user generated content easily, so it's always under attack of spammers. Mostly the aim of these cyber criminals is to steal the user's personal, professional, political, social or financial information by exposing the users with unwanted information on the web likes, pornography etc. in order to deceive them. There are number of methods by which the users' data can be hacked by these adversaries, and creating fake profiles to perform malicious activities on OSNs is one of the mostly employed methods. From the users' point of view, personal, professional and even financial data is no more secure. Figure 1 provides a quick view of various kinds of fake profiles and several other kinds of profiles found in different online social networks. Real profiles have to be categorised into compromised and non-compromised ones which are also shown in the figure 1.

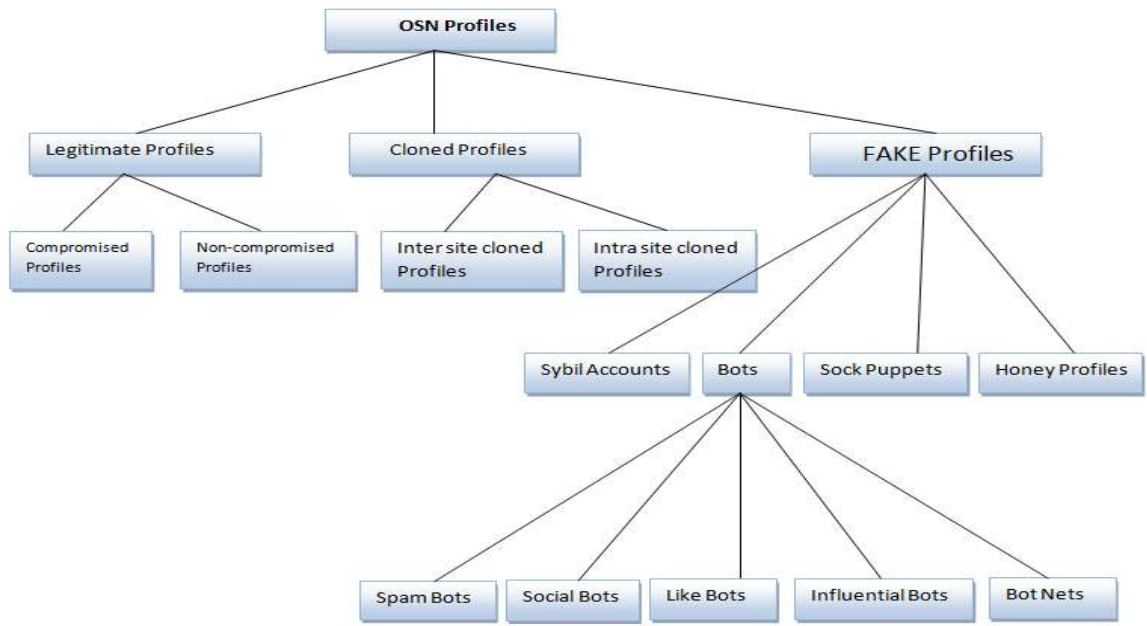

**Figure 1: Evolution of Fake Profiles in OSNs**

Profiles which follow the rules and regulations provided by particular OSN service are real. Here rules and regulations in the context of OSN may mean the owner should not have more than one personal account, it should not spread any unlawful, misleading, malicious, or discriminatory content, and it should not collect the user's information or access Facebook by automated means (such as bots and spiders)[10, 76]. A person handling more than one account; i.e. an account other than his principal account is categorized as fake[1]. Facebook has provided several options to enhance the privacy of user accounts like protecting the password and sending location specific login alerts and location alerts. Users can also use the extra security features of the network like how to logout from another device, how to keep Facebook password safe using app passwords etc. [87]. Moreover an approach called "Safebook" [85] is proposed in order to protect user's personal data from both the malicious users as well as service providers who violate privacy rules. In order to avoid cyber attacks, one should take proper care while using online social account. Also at the time of account creation, the terms and conditions should not be violated. This paper aims to put everything about online malicious accounts at one place. The remaining paper is organised as follows. Section 2 describes the different categories of profiles existing in online social networks. Section 3 explains several varieties of bots and their role as fake profiles. Different features to identify fake profiles are presented in section 4. Various data collection approaches have been explained in the section 5, section 6 presents different automated methods employed for fake profile detection. Finally, section 7 outlines pros and cons of several existing cyber laws to curb the online fake profiles.

---

[1] https://www.facebook.com/legal/terms

## 2. Online Social Network (OSN) Profiles

Based on the method of creation, targets and prevention mechanism, there are several flavours of profiles in OSNs. This section provides an exhaustive classification of different legitimate and phantom profiles with emphasis on online social networks.

### 2.1 Compromised Profiles

Compromised accounts are actually real accounts but their owners don't have complete control over these and they have lost the control to a phisher or any malware agent [7]. According to a study, [58] compromised accounts are the most difficult type of accounts to be detected. Another recent study says more than 97% profiles are compromised rather than fake [23]. The fake profiles are generally created to steal the credentials from the real users, and then fake profiles are abandoned or deactivated.

Compromised profiles have much value because they have already established a level of trust within their network and therefore cannot be easily detected and removed by the service providers. Attackers usually use compromised profiles with strict care in order to leverage the level of trust. The authors in [58], have presented an approach to detect compromised accounts from two popular online social networking sites, Facebook and Twitter by identifying profiles which show sudden changes in the behaviour by using statistical modelling and anomaly detection. Facebook has a system to recover hacked accounts once reported. There is an option "my account was hacked" on Facebook help page.

One more study [23] reveals that the compromised real profiles spread more malicious content than other types of fake profiles.

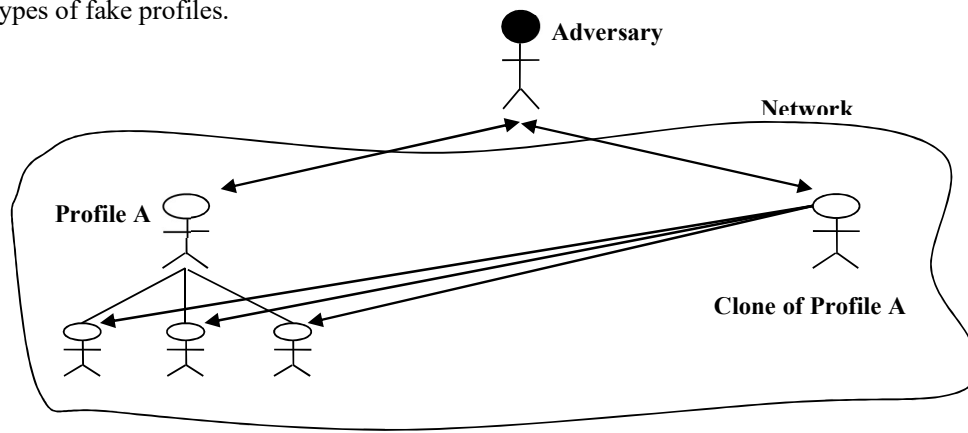

Figure 2: Intra site or same site profile cloning

### 2.2 Cloned Profiles

Profile cloning is the theft of identity from an existing user's profile and to create a new fake profile using stolen credentials. In other words we can say profile cloning is the process of stealing the victim's private information in order to create one more profile that can acquire the private information of victim's friends. These attacks are called as Identity Clone Attacks (ICAs) [24] which are of two types of profile cloning attacks namely single site and cross site profile cloning. The attackers are usually well funded, skilled persons and have almost everything available at their disposal and have control over compromised and infected accounts [27].The adversary can be a strange person, but statistics show that adversary has the knowledge of victim and can be one of victim's relative, friend or colleague [24].

The figure 2 and figure 3 depict the intra site and inter site profile cloning respectively which are explained below.

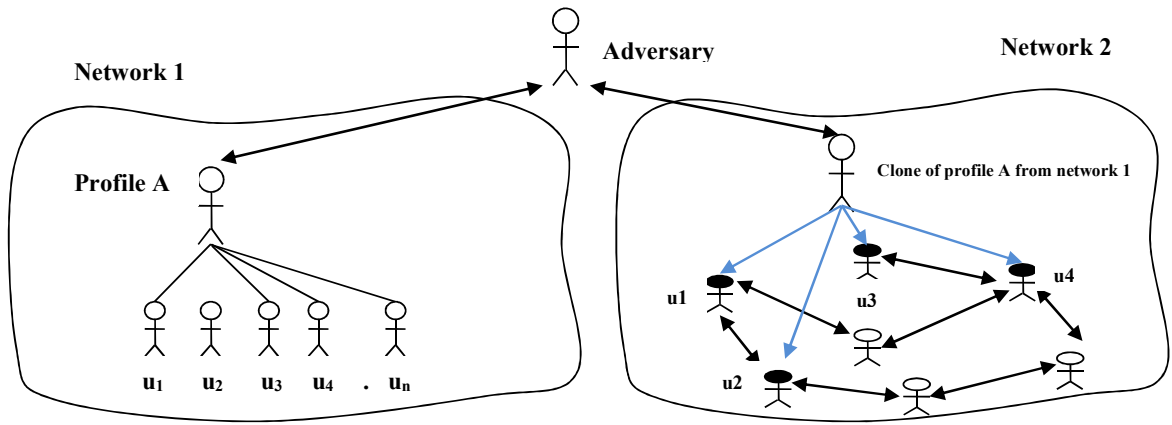

Figure 3: Inter site or Cross site Profile Cloning

**Intra site profile cloning:**

In case of intra site cloning, the adversary creates one more profile of the victim on the same network and sends the friend requests or follows the victim's friends. The victim's friends easily accept the friend request by treating it as a request from a legitimate user. It is possible to have different online accounts with the same name because in real life multiple persons can have the same name with different contact details, mobile number and email address. Adversaries are taking the advantage by creating one more account of an already existing person and pretending to be some real person with the same name.

**Inter site profile cloning:**

The inter site cloning (also called cross site cloning) is the process in which the adversary creates one more profile of victim in the new network where the user is not yet registered and sends the request to the victim's friends who are on both the networks. Inter site profile cloning can also be viewed as the reconstruction of victim's friend list in another social network where he/she is yet not registered. The main goal of adversary in creating these cloned profiles is to steal people's personal information, deceive or defame others or sometimes simply for entertainment. These ICAs are a matter of concern for both users as well as service providers as it becomes very difficult to detect these kinds of attacks. The users simply treat it as a friend request from a legitimate user and the service providers take it as a new user registering in these online social networks [6]. Detection of cloned profiles with more accuracy can enhance the level of security in OSNs which in turn will keep users safe from any kind of unauthorised access.

A recent study [28] has suggested different ways to cope with cloning attacks and recommended measures for OSN sites and users to improve the security. The authors in [6], have presented two automated ICAs namely 'profile cloning' and 'cross site profile cloning' and proposed prototype attack system (iCLONER) to attack the five most popular OSNs including XING, StudVZ, MeinVZ, Facebook and LinkedIn. This study showed that ICA schemes are much effective and enemies not raise much suspicion in users.

Profile cloning is a serious issue in online social networks. Normal users are not aware that their identities are being copied and used as a weapon to destroy their own kingdom by dodgy characters. These criminals actually copy all the content from your profile including profile name, profile picture education, work even status updates to give it exactly the same look as your real account. Then finally they report your real profile as fake by simply clicking on report option in Facebook[2].

Various approaches and techniques for the detection of cloned profiles are mentioned in the literature but still the profiles are cloned and misused at very higher rates. Behaviour of these accounts needs to be studied more rigorously to spot out their unusual pattern.

## 2.3 Fake Profiles

Fake profiles are different from cloned ones in many ways. In case of cloned profiles an adversary creates one more profile of the already existing one which is not the case for fake profiles. Cloned profiles are mostly created to extract the information of a victim or his/her friends whereas the fake profiles are used for various other purposes like spamming, advertising, etc. Some people create fake profile just to have one more account whereas some create multiple accounts deliberately to enter into people's sub graph. There are two ways to create fake profiles: one is created by writing a script and another is by manually creating one more account. And there are three main reasons for creating fake profiles: First, OSN service providers allow one account per mobile connection or per email-id, and to overcome this limit, people create one more account using different email-ids or phone numbers. Second is to enhance the popularity or the level of trust among the others. Third is to spread spam content among the real users. Fake entities exist everywhere on the internet like social networking websites, shopping sites, discussion blogs and forums, online dating websites, banking systems etc. [39] [40] [79]. And there is still need of strengthening security measures being employed by online social network sites to reduce cropping fake profiles and to avoid their hazards on social networks. Fake profiles are harmful for OSNs [33], [34] and can be more dangerous in future if not detected at early stage.

There is no doubt that researchers are rigorously focusing on the mitigation of fake online identities but the users' security and privacy is still not assured. Adversaries are too learning the strategies to hide themselves from the detection techniques. Creating phoney profiles in online social networks is an illegal activity and can lead to criminal charges and harassment[3]. There are several kinds of fake profiles in OSNs, some popular ones are described in the following subsections.

### 2.3.1 Sockpuppets:

Sockpuppet is an account developed with an aim to deceive others or to promote someone or something on discussion forums, blogs, social networking sites etc. In other words sockpuppets are those online accounts which are created to cheat the netizens in different ways for example to promote a particular product, convince people by saying there is low risk in an investment plan with high return [40] etc. Usually in case of OSN sites, the blocked users create new accounts which are referred as sockpuppetry [2]. According to the authors in [40], if there exist two different accounts on any news blog, social network or any discussion forum that belong to the same person, it is called sockpuppet pair.

### 2.3.2 Sybil Accounts

In case of Sybil accounts, the malicious users create multiple accounts and handle them manually to attack the trusted network. Sybil attackers have many goals like bad mouthing an opinion, in voting applications, to access resources, to compromise the security, and privacy [36] etc. In other words we can say when a node in online social network claims multiple roles and threatens the security; this is referred to as Sybil attack. For example in online voting system, a single user by using multiple IP addresses can submit a large number of votes, companies try to get popularity and higher ratings on Google page rank by using Sybil attacks, etc. According to [37], social networks with well-defined community structure are more exposed to these Sybil attacks because their links can be used by the Sybil attackers more effectively. Several studies [36, 37, 38 42, 68,77] have been carried out so far for the defence of these attacks, but still the detection of Sybil attacks is in its early stage. Most of the Sybil defence techniques work on ranking of nodes based on how well a node (an account) is connected to trusted nodes (legitimate accounts). A node has higher rank if it is within the local community of a trusted node. So in order to assign a rank to a node, most of the Sybil defence schemes focus on detection of local communities effectively. As soon as a user creates an OSN account, he/she becomes susceptible for targets of an adversary. Other fake profiles can catch one's behaviour and convince the user to perform unlawful activities. From above discussion, it can be concluded that fake profiles are basically of two types; one created manually and the others using automated methods. And automated fake profiles pose more threats than other kinds of fake profiles. Very recently, automated computer program called bot is being used to create automated fake profiles which are more harmful, the next section describes bots acting as fake profiles.

---

[2] https://www.facebook.com/help/306643639690823?helpref=uf_permalink
[3] http://www.cps.gov.uk/legal/a_to_c/communications_sent_via_social_media/#a10

## 3. Bots as Fake Profiles

A bot is a computer program that produces some data to interact with humans especially the persons using internet (netizens) in order to alter their behaviour [46]. More than 60% of the total web data is generated by bots [78]. Online bots also known as web robots or simply bot is a computer program that performs various tasks quickly and automatically which were not possible for a human alone. Basically the bots were designed to assist the humans to speed up their work and make it automatic. The main role of bots was to automatically aggregate contents from various news sources, work as an automatic responder to costumer queries, act as a medical expert to resolve health related issues and automatic travel guide. But nowadays the bots are misused by the public in various domains. In social networks, bots are used to retweet a post without verifying its source in order to make it viral. In online multiplayer games, bots are used to gain the unfair advantage [44, 46]. Sometimes bots acts as automated avatars to interact with humans and create social networks, which are even more difficult to identify [45]. Bots can also be used to influence users, posting messages and to send friend requests [47] in online social networks. The table 1 summarizes several types of suspected profiles and their distinguishing features highlighted in this section. Also the group of people/organization who are likely to get affected by the intrusion of these identities were also mentioned in the table. From the working point of view, bots are similar as Sybil accounts but the main difference is Sybil accounts are handled by users manually whereas bots are automated computer programs [46]. The main use of bots is web data crawling where a simple online computer program identify and extracts the information from web servers at much higher speed which was not possible by a human alone. Bots designed for malicious activities have become a serious threat for the internet. Various OSN service providers employed several ways to fight the spam bots. For example twitter and Facebook have added an option "report as spam" to identify a spam bot.

Facebook also has its Facebook Immune System (FIS) [27] to deal with such issues. But still the research in this domain is in its early stages. Users in various OSNs claim that their legitimate accounts are being caught by the detection techniques. According to a study [10], more than 8% bots exist in Twitter network. Most of them have been developed for commercial purposes. Bots can be of two types benign and malignant.

|  | Compromised Profiles | Cloned Profiles | Sock Puppets | Sybil Accounts | Bots |
|---|---|---|---|---|---|
| | Table1: Summary of various OSN accounts | | | | |
| **Definition** | Real user accounts whose owners have not the complete control over them and have lost the partial control to a phisher or any malware agent | Theft of identity of an existing user's profile and to create a new fake profile using those credentials. | Account developed with an aim to deceive others or to promote someone or something on discussion forums, blogs, social networking sites etc. | Multiple accounts manually created by malicious users to attack the trusted network. | Computer program that performs various tasks quickly and automatically which were not possible for humans alone. |
| **Purpose** | * To defame or deceive a person.<br><br>* To spread malicious content by exploiting the trusted network.<br><br>* To abuse legitimate accounts. | * Steal people's personal information<br><br>* Deceiving or Defaming a person<br><br>* Fun and Entertainment | * To honour, defend or support a person or an organization<br><br>* Manipulate a Public opinion<br><br>* To avoid a suspension or ban from a website. | * Bad mouthing an opinion, in voting application<br><br>*To access resources, and compromise the security, and privacy etc. | * To automatically aggregate contents from various news sources(data extraction)<br>* Work as an automatic responder to costumer queries<br>*Act as a medical expert to tackle health issues |
| **Effected Group** | Real account owners, user-friends. | Existing online users, People without online accounts, etc. | Bloggers, Wikipedia-users, Researchers | Netizens, Politicians, celebrities, Organizations etc. | OSN users, Bloggers, OSNs etc. |
| **Target Networks** | Facebook, Linkedin, Twitter, Online payment systems etc. | MySpace, Facebook, Linkedin, Twitter | Wikipedia, Facebook, Linkedin, Twitter | Facebook, Linkedin, Twitter | Facebook, Linkedin, Twitter |
| **Types** | Partial-Compromised(PC), Complete-Compromised(CC) | Intra site profile cloning, Inter site profile cloning | Strawman-sockpuppet, Meatpuppet | -- | Spam-bots, Social-bots, Like-bots and Influential-bots |
| **References** | [58][7] | [22][24][25][28][29] | [2] [39][40] [41][67] | [36][37][38][42][68][77] | [10, 44-53 ] |

Malignant bot designers may have many goals in their mind for example to change the mind setup of the person about a product, to spread and support fake or malicious news or to misdirect people [46]. Besides that there are various other uses of bots. Therefore, based on the functionality we divided the bots in five categories as shown in figure 4.

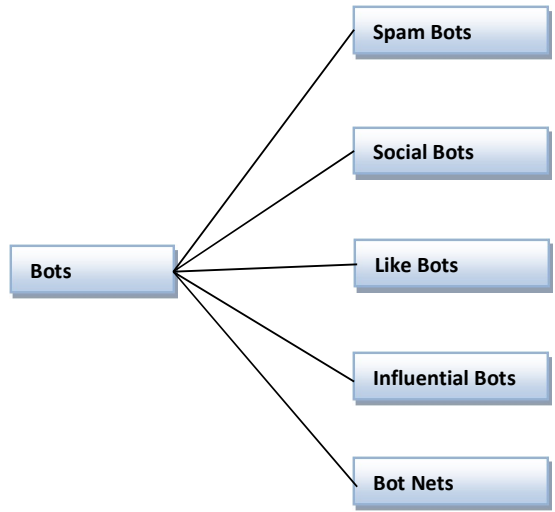

Figure 4: Categories of Bots in OSNs

Since all the bots have the same working mechanism and appear same to the normal people (not professional), based on the developer's intention and the context of online bots, we present their several categories in the following subsections.

### 3.1 Spam Bots

Spam bots are computer program which are designed for malicious activities only. They are designed with the aim to pollute the network by creating large number of unwanted relationships and spread the malicious content like links to personal blogs, paid contents, pornographic websites, and advertisements, or to shill for any person or organization, or to influence a particular article which is not that worthy. Spam bots are those bots which immensely spread unsolicited contents among users without their permissions [50] and are different from normal bots, which are developed for daily activities like weather update (e.g. Twitter bots) and social bots: which mimic a normal user. In the [48], the authors studied the behaviour of spam bots in twitter network and applied several classification techniques to differentiate them from normal bots. The authors in [48], have shown that out of four (decision tree, support vector machine, K-nearest neighbour and neural networks) classification techniques, the Bayesian classifier is best in predicting the class label (of a spam bot) on the bases of user's pattern. Although the spam bots are new to the OSNs but the detection of spams has been previously focused in the emails, web sites etc. A study [50] have developed a social bot with two components: an OSN profile and social bot software, and instructed it to perform operations like reading and writing the social content(spam), creating social interactions, behaving like real users and joining the online social communities.

### 3.2 Social Bots(or Sometimes Bots)

Social bots or simply bots are the computer programs used by humans for their several online activities. According to a study [49] social bots are highly complex computer programs which keep users busy and behave like humans. Bots are the programs which publicize themselves like viruses to reach and infect maximum number of users and exposed hosts [47]. According to the authors of [50] social bots are those bots which control accounts on online social networks and imitate the behaviour of legitimate users.

Social bots are not always problematic. They are same as other bots in their working but their focus is more on building social relations with the online people e.g. politicians can use social bots to get connected with their public, companies can use them as their customer care agents etc. Social bots imitate the human behaviour to gain the attention (for example followers, friend requests, replies, likes etc.) from their targets and use this trust network to spread content or promote an agenda or a product [49]. Also social bots play an important role in multiplayer online games to make the game more entertaining and interesting for the game lovers [44]. Sometimes social bots can also be used to gain the unfair advantages in the online games but here in this section we are talking only the positive aspects of social bots.

### 3.3 Like Bots

Like bots are just computer programs controlled mostly by advertising companies and politician to promote the products or to gain the reputation and trust in the network. Usually these bots are used just to like the content (unproven or verified) posted on social media but sometimes they post messages as well. In several OSNs one can buy fake likes for their content [51] and the sellers (usually cyber professionals) make use of multiple numbers of like-bots to like the customer content. Many application developers use the like bots to like there developed application on online stores. Also one (individual or organization) can purchase likes from various available online vendors[4]. The number of likes for a product or a page signifies its success and reputation. One of the main jobs of like bots is to click on the like ads or like pages. Advertising companies can use them for their own benefit but on the other hand can misdirect the normal users. Very few studies have been carried out for the detection of like bots (or fake likes) so far. In [51] a comparative study of 'likes' of Facebook pages produced by Facebook ads and several like farms. The authors created more than a dozen honeypot pages on Facebook and analysed the produced likes based on users' (likers') demography, temporal and social behaviour etc. Like farms can make the use of like-bots for their businesses, but naive users need to be aware about these fake likes, otherwise they can get unacceptable results. One more study [10] analysed a number of Facebook accounts used by some Like farms and compared their contents (posted on their timeline) with normal user content and found that Facebook accounts owned by like farms mostly produce likes and comments and most often produce the post the same content. Nowadays it is very hard to find that how many 'likes' for a product or a post are from real users and how many of them are fake which results in misdirection and poses false impact on the normal users. Therefore a strict mechanism should be developed and employed to identify the like bots and shut them off.

### 3.4 Influential Bots

Influential bots are automated identities that illegitimately perform discussions on some trending topics on OSNs like Facebook and Twitter in order to promote and popularize the topic [52]. Influential bots not only have the ability to escape from the detection techniques but can also obtain double followers than an average user and become influential in social communities.

Influential bots usually generate messages (tweets or posts) either by reposting (or retweeting) the content posted by other users on the same network or create their own synthetic message by a defined set of rules. Nodes (users) who are connected to the maximum number of nodes in the network are called core nodes and these core nodes play an important role in influencing a topic or an individual. Since the influential nodes have one of their aims to spread the content to the maximum number of people therefore they try to send maximum number of friend/connection requests before spreading the content. Influence of a particular node (user) depends upon its popularity and level of trust in the network. Popularity of an individual node is considered as the number of incoming requests or received messages [54]. Influential nodes play an important role in online marketing. Identifying influential nodes in a network often seems to be challenging for marketing companies. Therefore, nowadays the organizations first design their OSN bots and start getting into online communities to reach the maximum number of people. Once these bots obtain trust level within the real users in the network, they start promoting products or brands.

---

[4]https://boostlikes.com
 http://kingdomlikes.com

The main job of influential nodes in the network is to change the opinion of users about a particular topic or product. Influential bots in the same way try to change the way of thinking of users about an article or any brand on an OSN. Since the normal (real) influential users and the influential bots have almost the same job, therefore it is possible that they have some set of features in common. One possible way to identify the influential bots is to make use of tools and technologies like Klout[5] and Twitalyzer[6] which are used for normal influential identification. Various studies [55], [56], [57] have been carried out for the identification of influential users and influential bots in online social media.

**3.5 Botnets**

The network of automated computer programs in an OSN is referred as Botnet. Each program (bot) in this network is assigned either a similar or different set of tasks to be performed in an automated manner. A botnet is a collection of computer programs handled by a 'control-channel' which gives commands to perform unlawful activities [47]. Since the botnet consists of multiple bots, therefore the botnet controller can perform different kind of tasks like spreading malicious content (spam bot), promoting (liking) a post or a product (like bot), sending friend request to unknowns (interaction/social bot) and popularizing a topic (influence bot) at the same time.

Botnets are mostly controlled by malevolent users called 'botmasters' by issuing commands to perform malicious activities.
The primary goal of these botmasters is to extract the personal, professional and financial information of users on internet. Actually the main purpose of botnets was to assist the users in Internet Relay Chat (IRC) chat rooms [47] by controlling the interactions, providing help to administrators, offer games, extract information about the platform (operating system), and other details of the user such as email addresses, logins, aliases etc. In [10], the authors have studied the growth of social botnet in twitter network and observed how the tweets of a normal user differ from the content generated by social botnet and how these social botnets help in popularization.

Historically, botnets were primarily used to spread misinformation, propaganda and for many other malicious activities. Several kinds of bots get infiltrated into the target OSN to start a Botnet campaign. The botnet units (bots) help each other by liking the post without verifying it, influencing (retweet or share) the content of each other, writing the positive comment or review etc. in order to gain the trust in the targeted OSN. Botnets are mostly designed for different kinds of benefits varying from individual to individual e.g. shopping companies design them to get likes and increase the ratings of their products, researchers, academicians and data scientists use botnets to crawl data from the web, hackers and other cyber criminals use them as tools for social engineering. A study [53], defined a botnet as a set of bots (social bots in this case) controlled and supervised by a human controller called 'botherder' (enemy), the authors of this study designed the botnet with three components namely socialbots, botmaster and control-and-command-channel which handles the targeted OSN profiles, providing commands (like posting a message, sending friend/connection request etc.) and carrying the commands respectively. These botnets were designed to extract the data from the internet.

---
[5] https://klout.com
[6] http://www.twitalyzer.com

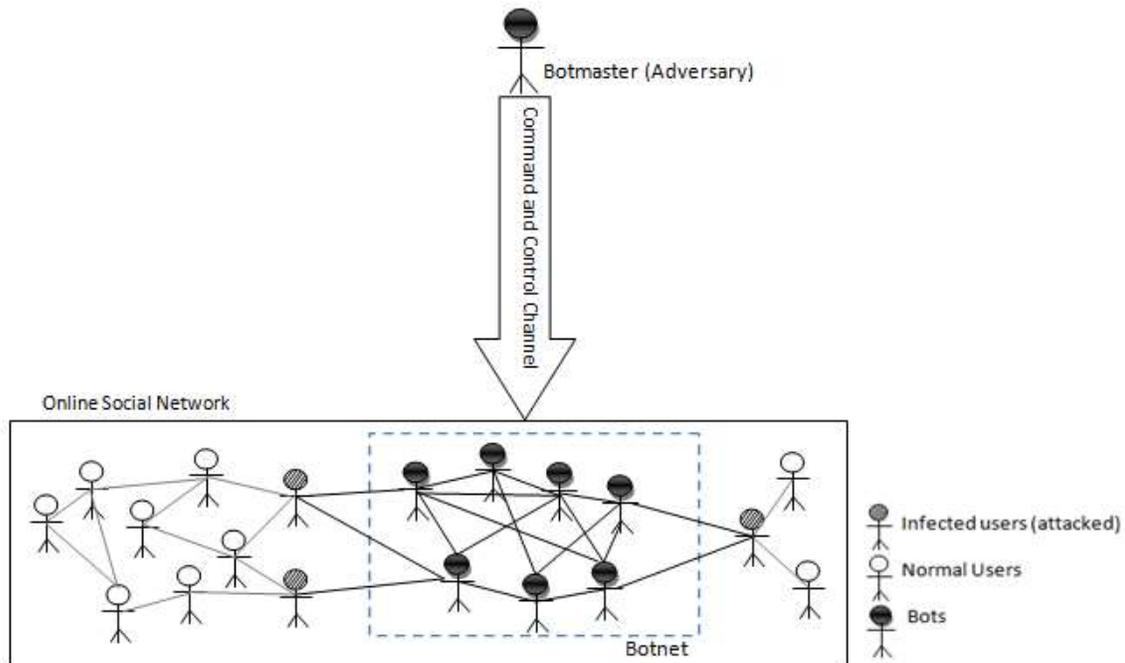

Figure 5: Pictorial representation of an OSN Botnet

A pictorial representation of a typical botnet is shown in the figure 5. There are three types of users in the diagram viz. normal users, infected users and bots. Botmaster is simply the user (adversary) who owns and controls the botnet, and provides the commands via command and control channel. The commands are followed by each bot in the botnet. Botnets are very complex and highly evolving threats to users' trust and security on the internet.

Therefore serious strategies and steps should be taken to mitigate their effect and risk associated with them. Since OSNs are for real humans, handling profiles by automated programs is against the rules and regulations of OSNs. Bots, either used for commercial activities, entertainment or research must obey the cyber rules and regulations. Table 2 gives a quick view of some of the characteristics of bots and botnet. Bots use OSNs as an attractive medium to spread the abusive content, bias public opinion, influence user perception and perform fraudulent activities etc.

These bots may be used by people based on their requirements. In the table 2 we have also shown that which category of people may use which type of bot. Nowadays we can see the reviews for the different products on the shopping websites, some of these are not from the real accounts. Many shopping sites make use of bots to generate the positive reviews for their product. Sometime we call them shopping bots. As reflected in the table, politicians and celebrities usually make use of influential bots to raise their craze in the audience. Most of the researchers and academicians make use of social bots to create social relations with public and extract specific information from them. These bots were also created by dodgy characters to perform malicious activities.

As soon as one creates an OSN account, he/she becomes susceptible for targets of an adversary. Other fake profiles can catch one's behaviour and convince the user to perform unlawful activities. From

| Table 2: Bots and their specific features | | | | | |
|---|---|---|---|---|---|
| | **Social Bots** | **Spam Bots** | **Like Bots** | **Influential Bots** | **Botnets** |
| **Purpose** | Create social relations | Spread malicious content | Increase the ranking/ratings | Change the behaviour of people, to increase popularity | Campaigning of bots |
| **Used by** | Politicians, Researchers, academicians etc. | Cyber criminals, Advertising agencies. Fake Agents | Advertising agencies. | Politician, Advertising companies, celebrities | Researchers, Hackers, Advertising companies etc. |
| **Networks** | Online Social interaction networks, Blogs, Discussion forums | OSNs, online shopping sites etc. | Shopping websites, OSNs, | OSNs, discussion forums etc. | All types of OSNs |
| **References** | [19, 27, 44-47, 49, 50, 52] | [50,48] | [51,52] | [50, 36, 34] | [47,52, 53] |

above discussion, it can be concluded that fake profiles are basically of two types; one created manually and the others using automated methods. And automatic fake profiles pose more threats than other kinds of fake profiles. A botmaster can handle several fake profiles simultaneously (botnet) which can damages the reputation of network to great extent. Therefore, in order to assure the privacy and security of user data and reputation of the network, automated fake profiles should be detected and removed from the network.

## 4. Feature Selection for Identification of Fake Profiles

Fake profiles in online social networks (OSNs) are becoming a serious issue day by day. New techniques and strategies are employed by cyber criminals to escape from detection techniques or to totally bypass them which put genuine users' personal, professional as well as financial information at high risk.

Researchers and developers are continuously putting forward their effort to mitigate the effects of fake profiles on netizens and OSN service providers. Different features of online profiles have been identified by several researchers from time to time to train their detection models [2,8,31,32,35,39,40,41,44]. Since the adversaries are also aware of these detection techniques, they also keep changing their behaviour according to the situation. Therefore some fake profiles always exist in the OSNs.

Different attributes identified and/or employed by various studies for the detection and identification of fake users in OSNs are presented in the following subsections. Based on the characteristics and the behaviour, user attributes may be categorised in two broad categories as shown in the figure 6.

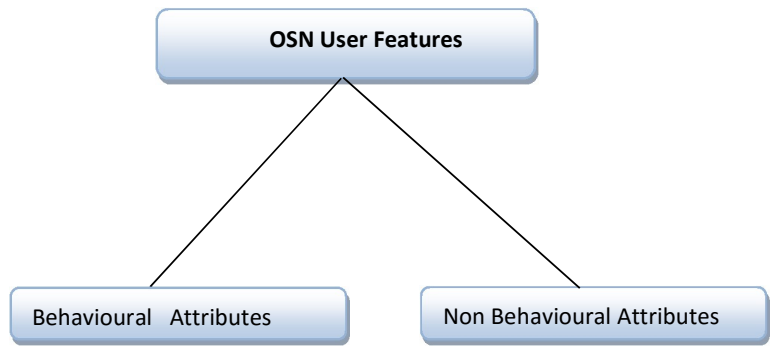

Figure 6: Categories of User attributes

### 4.1 Network Based Attributes

As in real life, people interact with friends, discuss some issues, make new friends etc., the OSN users also perform their daily online social activities like following news updates, interaction with online friends, making new friends, and joining communities (groups, pages, events etc.) In this way they create a network of trust with their friends. Various researchers exploit this network of trust and apply several statistical measures to extract various informative attributes called as network attributes or network features. For example number of communities (pages, groups, events) a user belongs to, in-degree (friend requests received), out degree (friend requests sent)[7], betweeness centrality, closeness centrality etc. Based on these network attributes various studies put forward their models for the detection of fake profiles in OSNs. Some attributes reflect the actual behaviour of a user while some features reflect their non-behavioural activities. In [35], the authors have exploited the behavioural feature like 'pages liked by the user' and how likely the user likes pages (Rate of like activity) for the detection of analogous user behaviour in the Facebook network. If the frequency of page likes within a specific time interval is very high or all the pages liked by a user strictly belong to a single category, the user may be considered as suspicious. In [44], a machine learning based approach has been used for the detection of social bots in twitter network. The author has used non behavioural features like number of friends, number of followers and followee ratio of a user profile. A study on Wikipedia users has used verbal behaviour like punctuation mark and use of lowercase or capital letters to detect the sockpuppet accounts [41]. In [8], the authors have used non behavioural features of SinaWeibo[8] users to detect spammers on the network.

The non behavioural features which are generally incorporated by various authors/data scientists include number of reposts, number of comments, number of likes, number of mentions, number of urls in the post and number of hashtags. These features can be further categorised into profile and non-profile based features. Age, location, gender, profile image etc. can be considered as profile attributes. Nature of the posts, active friends etc. are categorised under non profile attributes. Both of these are highly exploited by researchers to identify spam on OSNs. Furthermore according to twitter spam policy, if the number of people following you is less than the number of people followed by you, or you trying to follow the people beyond the limit[9], your identity can be considered as suspicious. So, this network feature is a recommended feature for the identification of fake profiles in twitter network.

---

[7]*Friend request sent* and *friend request received* are Facebook terminologies.
[8]SinaWeibo is a Chinese micro blogging website www.weibo.com
[9] https://support.twitter.com/articles/68916

## 4.2 Content based attributes

What a user posts, sends or shares on his/her profile is the content that tells a lot about the behaviour of that user. Not only behaviour, person's way of thinking, even his/her overall personality on the network is reflected by his/her content. In the literature, researchers have exploited several content based attributes to identify various kinds of spammers on several networks. Anything that is uploaded on a user's profile can be considered as content. Number of messages posted by a user, number of tags in the post, number of words in the post, number of pictures uploaded, number of posts shared etc. all are content based attributes. In [41] authors exploited content based features like use of capital or lowercase, quotation count and punctuation mark for the detection of sockpuppets in Wikipedia. Another study [8] on a Chinese social network SinaWeibo uses attributes like number of reposts, number of comments, number of likes, number of hash tags and urls in the post in order to detect the spammers on the network. Similarly the content based attributes like number of posts by the person, number of photos of a person tagged in, the number photos the person has uploaded and number of tags in the uploaded photos by the person have been used for the detection of fake profiles in Facebook social network [21].

At the initial stages, researchers were more focused towards the url centric features to detect human controlled fake profiles. But nowadays social bots have replaced these human adversaries and have become very active. Besides the urls in the post, the malicious text, pornographic pictures etc. can also be identified by using content based characteristics. Therefore, the opinion oriented (content based) features have been exploited by various researchers for the identification of spammers in the social networks. Apart from above attributes, there are other attributes used in the literature like POS (Part Of Speech) tags, n-grams (unigram and bigram). In order to understand the content and meaning of posts the authors in [9] have done a case study on fake identities in social media and have listed several features of a Facebook profile to conduct a social engineering experiment on Facebook. Mixed attributes have been used by the authors for their experiment. As per the study of authors in [59], if any account posted a duplicate content, it can be considered spam account because genuine accounts do not update same content multiple times.

Different researchers use their different terminology to represent feature sets of an OSN profile, some call them graph-based features, while some use the term neighbour based features for them.

## 4.3 Temporal Features

Temporal features as the name implies are the characteristics related to time. e.g. account creation time, last login time, time between the two status updates, active account time [58] etc. According to [58], if the users are posting messages or showing any other activity on their profile in quite odd periods (regular sleeping hours), that is considered as anomaly. In case of social bots, a group of accounts (botnet) are getting active at the same time, perform some activities (usually malicious activities), and logout at the same time. Because socialbots are controlled by a single adversary, this behaviour can help researchers to identify a social botnet.

## 4.4 Other features

Besides above category of features, literature is replete with other profile characteristics which can be used to identify anomalies in several OSNs e.g. languages known by the account owner. Usually the user is free to post in any language, but it is expected that the user may know only a few languages. This can also help in identifying the anomalies in online social networks. In certain scenarios some users mention multiple languages in their "languages-known" column later they write posts in some other language which is not the normal behaviour. Authors in [58] have used a library, libtextcat to determine the language in which the message was written. This library performs language categorization based on n-grams. Another feature can be the message title. Usually the users have a set of their topics about which they are generally carrying out discussions like their favourite sports, favourite movie or some political discussion. Since, the users typically post about their favourite topics and then suddenly post about some unrelated subjects and topics, this unusual behaviour can also help to spot an anomalous user. Natural language processing (NLP) techniques can be used to detect the topic of the message. Twitter has the "hash tag" mechanism by which we can say what a particular post is about. Furthermore, a user posting duplicate tweets with different @usernames from the same account can be suspicious [59]. In one more study [5], towards the characterization of real profiles, authors have identified three features including growth of OSN friends with time, authentic social interactions and change in structure of OSN graph over time. Under the third feature, average degree of nodes and number of singleton friends are taken into consideration to detect the fake profiles. Different users have different set of features, specific to the network they are associated with. Therefore, to spot out the malicious users in a specific social network, a study of the environment and the culture of that network would be helpful. Table 3 shows the several set of features incorporated by various researchers to identify the malicious users in various online social networks.

As summarized in the table 3, we have categorised the attributes in two categories namely behavioural and non-behavioural. These two categories can be further divided into network, temporal and content based attributes. This exhaustive list of features and their categories identified, would certainly be helpful for researchers to perform different kinds of research in OSNs.

**5. Data Collection Approaches**

Data is the primary need in any analysis. Since several online social network service providers have imposed the restrictions on automated crawling, and in several other OSNs only few publicly available attributes can be extracted. Therefore, the main challenge in analysis of online social networks is to obtain the updated data and that too in real time.

But still there are various ways to obtain the data for the research purposes. In the following subsections we will highlight some of the data collection approaches used by various researchers, which are shown in figure 7. This figure highlights four (Honey profile based, API based, Bot based and Artificial data generator based) popular and commonly used data collection approaches to extract information from the social networks.

**5.1 Honey Profile based Approach**

Honey profile as the name indicates are the online profiles used to attract the people towards them. These kinds of profiles were created by researchers, academicians and normal public to extract the data, promote the products and brands or used simply for entertainment. Honey profiles are created and designed according to the need and the situation. In [3], the authors have created a set of honey profiles in three largest social networks to collect the data of spam activities. Honey profiles or simply honey-traps are of several variants, some people create honey profiles which attract teenagers and young people on the targeted network and leave them for some time, while some people create honey profiles to attract general public and therefore they repeatedly keep on updating these profiles with latest updates and interesting stories and images.

| Table 3: Categories of features used by various researchers for Malicious Profile Identification ||||| 
|---|---|---|---|---|
| **Features/Attributes** | **Author(s)** | **Network** | **Type of Feature** | **Purpose** |
| - Profile layout colours<br>- First names, user names<br>- Spatiotemporal Information | [31] | Twitter | Behavioural | Detection of spam bots |
| - Pages liked by the user<br>- Rate of like activity | [35] | Facebook | Behavioural | Anomalous User behaviour detection |
| -Writing style etc. | [39] | Tianya (China forum) | Behavioural | Sockpuppet Detection |
| -Number of Replies<br>-Registration Dates | [40] | Uwants (Hong Kong discussion forum) | Behavioural | Sockpuppet Detection |
| - Ratio of friend requests sent to the number of friends<br>- Ratio of messages containing url to the total number of messages,<br>- Similarity among the messages sent by the user,<br>- Message sent,<br>- Friend Count | [32] | Xing | Behavioural | Multiple account detection |
| - Number of total revisions (Rt)<br>- Article discussion (Rdt)<br>- User page (Rut)<br>- User discussion page (Rtt), | [2] | wikipedia | Non-Verbal Behaviour | Multiple account identity deception |
| -Punctuation count<br>-Quotation count<br>-Use of capital or lowercase | [41] | wikipedia | Verbal attributes | Sockpuppet Detection |
| - Number of friends<br>- Number of followers<br>- Follower ratio | [44] | Twitter | Non Behavioural | Detection of spam bots |
| - Number of reposts<br>- Number of Comments<br>- Number of Likes<br>- Number of Mentions<br>- Number of urls in the post<br>- Number of Hashtags | [8] | SinaWeibo | Non Behavioural Social Behaviour | Spammer Detection |
| - Educaion and work<br>- Relationship status<br>- Gender<br>- Number of wall posts by the person<br>- Number of photos of person tagged in | [21] | Facebook | Non behavioural | |
| - Number of photos the person has uploaded<br>- Number of tags in the uploaded photos by the person | | | | Detection of Fake profiles |
| - Microblogs<br>- Followers<br>- Followings<br>- Friend Number<br>- Number of microblogs to get a fan | [12] | SinaWeibo and TancentWeibo | Non Behavioural | Analysis of Spammers |

The authors in [3] have created more than 890 honey profiles on three different social networks Facebook, MySpace and Twitter to gather the users' information. Similarly the authors in [9] created eight very interactive and attractive profiles on Facebook with different age groups to collect the user information; the authors have observed these eight profiles for a period of three months.

Another study [51] has used honey profiles to uncover social spammers in social networks. As the honey pots attract users' attention on the large scale therefore spammers make the use of these profiles to spread malicious activities as well.

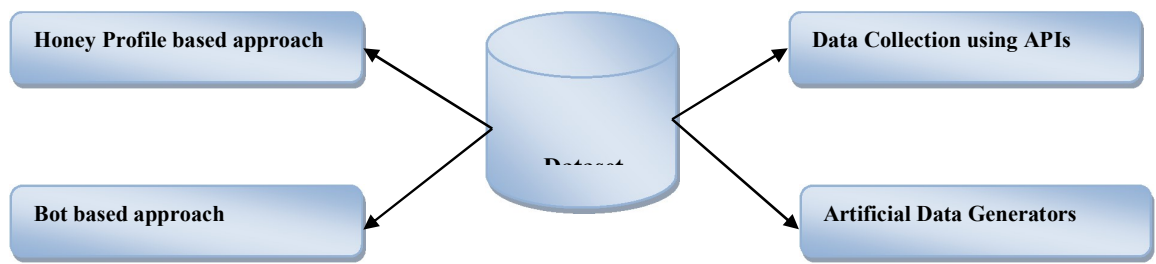

Figure 7: Data Collection Approaches

### 5.2 Data Collection using APIs

Collecting data using APIs (Application Programming Interfaces) is mostly used in almost every social network analysis and is highly recommended. Generally the OSN service providers assist developers and normal users with the several libraries (packages) for various data extracting activities. In order to collect the work-specific data from these social networks most of the researchers write their own code using APIs to interact with the targeted services. Almost every social networking site has its own API for example there is GRAPH API[10] for Facebook network which allows its users to interact with their application and collect user information. Similarly for Twitter there is Twitter API[11]. Some researchers and scientists design their own data crawlers to extract data specific to their research from social networks. A study in [4] has extracted data from Facebook and Twitter networks in order to detect spams in these two social networks. In [3], the authors have written a script to get connected with already created honey profiles and extracted all the information needed to detect the malicious activities. In [60], the Twitter API methods have been used to crawl the activity of users, their 100 most recent tweets and their following/followers. The authors in [14] have used java API "HTML Parser" to collect the users' public information from Facebook network. A similar approach was used in [5], authors considered Facebook as a target OSN and developed a Facebook sensing application to collect the required statistical information from the profiles. The researcher in [61] has implemented an application called "NetVizz" for the collection and extraction of information about Facebook users. Similarly there are various other studies in the literature that make use of these APIs for user data collection. Figure 8 gives a conceptual view of typical data crawler used to extract the data from social networks.

---

[10]https://developers.facebook.com/docs/graph-api
[11]https://dev.twitter.com/overview/api

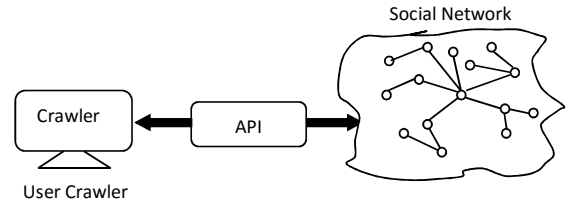

Figure 8: A typical OSN data crawler

Crawler is simply an application which interacts with the service via API. Crawlers mainly can be BFS (Breadth First Search) or DFS (Depth First Search) based. In case of BFS crawlers, the data collection process starts from a target profile, its immediate neighbours are discovered and their public fields are extracted. The authors in [19, 61] have developed OSN data collector based on BFS algorithm. In case of DFS crawler, instead of extracting all the neighbours of the target profile, the attributes of neighbour's of the neighbour profile up to a specific level are collected first. Then immediate predecessors are explored and their fields are recorded. One of the main advantages of DFS method over BFS method is that it has lower memory requirement. But in order to deal with large social networks BFS based crawlers are the good choice. We can also select the profiles at random for data collection, but if we process them in a sequence we can have many hidden patterns and observations. One more effort [65] towards the collection of Facebook data, has designed a data crawler based on simulation interaction and eliminated the shortcomings of a previous study [19] which was limited to 400 profiles only. Besides highlighting the privacy concerns of users, the authors, also showed that more than 35% Facebook users have altered the default privacy setting for their accounts.

Once the data is collected, labelling is done manually in many cases by crowdsourcing. Crowdsourcing is the process of manually generating data by a number of paid people, on the internet. Amazon Mechanical Turk (MTurk)[12] is a crowdsourcing Internet marketplace used by business organizations to obtain the human (aka turkers) generated data.

Manual exploration of profiles is carried out by human experts [14]. These labelled records are then used as training data with several machine learning techniques.

### 5.3 Artificial Data Generators

Since in many cases the data is not available or the OSN service providers have put restrictions on auto crawling to ensure the security and privacy of the users' data, therefore we need to generate the artificial data to test algorithms and tools for SNA. The data can be generated using various available tools based on the known statistics or parameters of any existing social network. For example if we know the degree distribution, clustering coefficient, average betweeness centrality and other statistical parameters, a dummy data set can be generated for analysis purposes. In order to represent an authentic data, the synthetic data are generated [64] and are helpful to design systems and to fulfil specific requirements which may not be possible with original or real data. Various online data generators (e.g. GEDIS Studio [110], Databene Benerator [111]) are available for the generation of artificial data[13].

---

[12] https://www.mturk.com/mturk/welcome
[13] http://www.freedatagenerator.com, http://www.generatedata.com, https://www.mockaroo.com

Artificial data can have limitations too, for example the behaviour and the statistics of the network may change with the time. Therefore the results obtained from the synthetic data may not always be correct and accurate according to the current statistics of the specific network. But if we do not have original data available at the time of testing, the synthetic data is the best alternative.

The fourth approach for the collection of OSN data is the Bot based approach. This is one of the most popular and commonly known approaches for collection of OSN data. In this method, people write computer program (crawlers) using several programming languages like java, python etc. to interact with APIs in order to crawl the social networks and extract the required information. This automated computer program is known as bot. Section 3 described several kinds of bots found in most of the OSNs.

**5.4 Some well known data sets**

Researchers and developers in most of the cases make use of APIs provided by OSN service providers to collect the required data. But in some cases, where data is much in volume or some constraints on auto crawling mechanism were imposed by service providers, the researchers attempt to write their own script using different programming languages to extract the required data. Still several attributes (mostly private) are inaccessible and cannot be extracted.

| | Table 4 : Popular Data Collection Approaches | | | |
|---|---|---|---|---|
| | **Honey Profile Based** | **Using APIs** | **Artificial Data Generators** | **Bot based approach** |
| **Description** | Collection of data using attractive fake profiles | Allow developers to read from and write data into OSN in an efficient way | Generation and Collection of data using existing tools and techniques based on known statistics of OSNs. | Use of automated computer programs for the collection of user data from online social networks. |
| **Mostly used By** | * Researchers<br>* Academicians<br>* Organizations<br>* General people | * Developers<br>* Security<br>* Organizations and labs | * Students<br>* Academicians | * Developers<br>* Researchers<br>* Academicians<br>* Organizations |
| **Reference** | [3, 30, 69] | [1, 5, 9, 14] | [26, 64, 62] | [46,47,50] |

Various Datasets are publicly available [70], [72], [74] while some need a registration process or request for the same via email [71], [73], [75]. Table 4 provides a quick glimpse about several data collection approaches available in the literature.

In early time of social networks, there was no restriction on the crawling of user data. Most of the user features were easily available but at that time the research was not strictly focused towards online OSNs. Presently, the researchers are paying great attention towards this domain, but unfortunately there are restrictions imposed by service providers on data crawling pertaining to more stringent security and privacy of different OSNs. Only a limited data is available and accessible which hardly contributes to any sound research work. Several automated methods are used by various researchers for data collection as summarized in this section; still more efficient methods are required for better research.

## 6. Automated methods used for fake profile detection

In section 2 & 3, several types of OSN profiles and their properties have been described. In the present section, a number of techniques used by various researchers to detect fake profiles have been presented. In [25], authors have divided the Facebook network into smaller communities based on their similarities, all the profiles similar to the real profiles are gathered to calculate the strength of relationship in order to check whether it is a clone or not. One more study [28] has presented an approach to stop and avoid the cloning attack by use of an attacking methodology, Fake content is injected into the network and an ICA is carried out to collect information during a conversation between the clone and real profiles. In another study [29], authors have propose a method for detecting social network profile cloning by designing a system with three components namely *information distiller, profile hunter and profile verifier*. Information distiller extracts the information from real user profiles and selects attributes which can be used to uniquely identify the profile. Profile hunter processes the information passed by information distiller and locates the profiles of the user on different OSNs to create a profile record which contains a link to the user's real profile and links to all the profiles returned by the result. Profile verifier calculates the similarity score between all the profiles and presents the result to the user.

For fake profile detection various techniques have been employed. The authors in [30] have used Markov CLustering (MCL) technique on real data of Facebook network for the identification of fake profiles, the authors have identified a set of features related to Facebook fan pages, links shared and active friends to model a social network weighted graph and applied MCL to categorize the different types of accounts. In this graph the node represents a profile and an edge represents a connection between two profiles. Given a matrix of nodes as input, MCL groups them into three clusters one of which contains all the fake identities, second contains all the normal profiles and the third cluster contains the mixture of both. For the third group of profiles the majority voting concept is applied to identify the class of outliers and merge them with an appropriate cluster. This study also suggested that techniques like decision tree, Support Vector Machine (SVM), Naïve Bayes (NB) etc. can be a choice to classify profiles as fake or real, but do not work efficiently for the social network profile dataset with multiple classes. Also there is scarcity of such well defined profile data sets and most of the data sets neither have any predefined class label nor a well-defined feature set, therefore unsupervised learning techniques are preferable over supervised techniques.

Similarly a novel approach has been proposed by authors in [31] for deception detection in Twitter social network using gender and location attributes. Bayesian classifier and k-means clustering were applied on gender specific attributes like profile layout colours, first names, user names and spatio-temporal information to analyze the user's behaviour. Another effort that tracks cyber criminals who own fake online accounts, is based on the concept of honey profiles. Authors in [3] have created a set of honey profiles on three popular OSNs, collected the data and identified the anomalous behaviour of users who got connected with the honey profiles. A tool has been developed based on the identified features. Using the spam strategy, four categories of spam profiles were distinguished, which are: Displayer, Bragger, Poster, and Whisperer. The study has focused only on the identification of Bragger (who post the message on their own walls so that it will reach to all the victims) and Poster (who send direct public messages to the each victim) spammers. Six features including FF ratio (R)-ratio of friend requests send to the number of friends, URL ratio (U)-ratio of messages containing url to the total number of messages, Message Similarity (S) - similarity among the messages send by the user, Friend choice (F), Message sent (M) and Friend Number (FN) were extracted. And finally machine learning techniques were applied to spot fraud and real profiles on Facebook, twitter and MySpace.

A study [2] based on computational approach to detect multiple accounts of an individual is presented given his nonverbal behaviour using Wikipedia for experiments. The authors also demonstrated how developers and designers can use non-verbal features for the detection of multiple accounts (belonging to the same person) to keep their online social networks safe. The authors applied SVM, Random Forest (RF) and Adaptive Boosting Algorithm (ADA) on publicly available datasets to show the efficiency of the method.

Various other researchers put their successful efforts for the detection of sockpuppets. A study in [39] proposed an algorithm based on combination of authorship identification technique and link analysis. The link between two nodes is created if they have the same interest in most of the topics and their set of comments are extracted to check their writing style and a hypothesis is created that these two sets belong to the same person and t-test is applied. If null hypothesis is true, the edge between two nodes is maintained otherwise eliminated. Link-based community detection technique has been used to reduce the network. Early sockpuppet detection is used to prevent online frauds like content intimation, theft of user's info and cyber cheating etc. Another method to detect sockpuppets on a Hong Kong based discussion forum is presented in [40]. The method is based on the total number of topics posted by one account and the number of replies from the other accounts. A detection score is calculated and if it is larger than the threshold then there exists a sockpuppet pair. The larger the score, the more will be the chances of two accounts being the sockpuppet pair. Based on the verbal features like of the users Punctuation count, Quotation count, Use of capital or lowercase, the authors in [41] presented a sockpuppet detection method for Wikipedia network using natural language processing techniques.

Researchers also paid a vital attention towards the detection and defence of Sybil accounts and their attacks respectively [36, 37, 38 42, 68, 77], but the detection of Sybil attacks is still in its infancy. Most of the Sybil defence techniques work on ranking of nodes based on how well a node (an account) is connected to trusted nodes (legitimate accounts), the node has higher rank if it is within the local community of a trusted node. Therefore most of the Sybil defence schemes focus on detection of local communities effectively in order to assign ranks to the nodes. In SybilGuard [38] the authors present a novel approach to protect a social network from Sybil attacks. They consider the link between two nodes as a trust relationship. Sybil nodes are differentiated from trust nodes using the calculated trust-relationship. Actually SybilGuard depend upon two characteristics of underlying social networks, first, the trusted accounts always have huge number of links, second, the fake users
create many nodes (accounts) but with few attack edges. According to [42], one of the approaches to prevent Sybil attacks is to have identities which are certified by the trusted agencies; otherwise we are likely to get unacceptable results. There are various agencies for the authentication of nodes for example CFS cooperative storage system[43] authenticates each node by its bunch of IP addresses and the EMBASSY assigns each machine a set of cryptographic keys and embeds them in device hardware.

Bot detection is also taken into consideration by several researchers. The authors in [48] have shown that out of four (decision tree, support vector machine, K-nearest neighbour and neural networks) classification techniques, Bayesian classifier is the best in predicting spam bots in Twitter network. Various studies have been carried out for the identification of influential users (bots) in online social media like [55], [56], [57]. In [10] the authors have studied the growth of social botnet in twitter network and observed how the tweets of a normal user differ from the content generated by social botnet and how these social botnets help in popularization. Table 5 provides a quick glimpse about several techniques used for fake profile detection in various online social networks.

As evident from table 5, almost all kinds of machine learning techniques have been used for detecting malicious profiles. The content generated by OSN grows exponentially. Also the number of profiles being added to different OSNs is growing exponentially. The authors conducting any kind of research on OSNs basically need to handle big data. Big data handling techniques must be scalable as big data cannot be processed or stored on a single machine rather it is to be distributed over a cluster. So, one needs to apply scalable machine learning algorithms such as large scale graph analysis, graph partitioning, and clustering algorithms for fake profile detection in OSNs.

## 7. Cyber Law and the Fake Profiles

The growth of social media has comforted people to create and share content about their day today activities, with the individuals from different corners of the world freely and without cost. The wildfire growth of social media era is silently breeding several relatively new legal and social issues across the web. The increased global use of social media has facilitated a number of new ways to commit crime. For example, profile pictures and messages updated on social media sites like Twitter, Facebook etc. are being used by cyber criminals to carry out unlawful operations. With limited government oversight and inadvertent users, more and more social media members are exposed to cybercrimes. Cybercrime is a crime to perform illegal activities using computer as a tool, like identity stealing, pornography, committing fraud using somebody's intellectual property, or violating privacy. Cybercrime, exploiting the Internet, has attracted the attention of researchers as the computer has become central to commerce, entertainment, and government. The Internet is not capable enough to limit the geographical and jurisdictional boundaries, but physically the netizens remain in jurisdictions and are subject to laws, independent of their presence on the Internet. A single transaction may involve the laws of at least three jurisdictions: the place of the user's residence, the place where the server is located and state or nation where the transaction takes place [92]. The term 'cyber crime' is a misnomer. There is not any definition for this term in any law endorsed by the National Parliament of any country. Basically the cyber crime is same as the concept of conventional crime. Both cause violation of rules of law and counterbalanced by the sanction of the state [93].

Basically cybercrimes has 3 major categories namely Cybercrime against an individual, property and Government. Cybercrimes against persons include various crimes for example Identity theft, dissemination of obscene materials or pornography, defamation, harassment with the use of a computer such as hacking, cracking and cyber-stalking. While all countries generally have laws prohibiting these cybercrimes and have their penalties typically varying from country to country. Identity theft is quite prevalent and is among the fastest growing criminal offenses in the world. According to a report by U.S. Bureau of Justice Statistics (BJS)[14] more than one million American citizens had their identities fraudulently used to open bank accounts or credit card accounts. Also according to the report more than 16 million Americans were victimized by account theft by use of stolen ATM and credit cards [96]. In the case of online transactions, criminals use "website cloning," whereby a financial institution's Web site is duplicated or "cloned" in order to contact unsuspecting credit card customers who then unknowingly provide personal information to be used to access their legitimate accounts. Cyber defamation with the involvement of any virtual medium is same as conventional defamation [94].

Unauthorized access over computer system is commonly referred as Hacking .Which is a dreadful feeling to know that someone has badly sneaked into your computer systems without your consent and has corrupted the confidential data and information [95]. According to the Information Technology Act 2000[15] existing in India, hacking itself is not a crime but looks into factor of mens rea. As per the 66 (b) section of the Information Technology Act 2000, there is punishment of imprisonment for 3 years and fine which may extent to two hundred thousand rupees, or both [97].

---

[14] https://www.britannica.com/topic/cybercrime/Identity-theft-and-invasion-of-privacy
[15] http://www.legalserviceindia.com/article/l146-Cyber-Crime-And-Law.html

Cyber harassment is the crime of violating the privacy of netizens. This privacy violation is a very serious nature of Cybercrime. Nobody likes any other person to sneak into his or her own privacy which is granted by internet to its users [98] but inadvertently users are themselves exposing their privacy by simply logging in their facebook or Google account and making their location public. In the United Kingdom, several steps have been taken with regard to the issue of Internet privacy. The Data Protection Act, 1998 was enacted [99] on July 16, 1998, came into force on March 2000 for implementing the European Union's Data Protection Directive. This is one of the most important cyber laws for protecting Internet privacy in Great Britain. This form of legislating has been backed up by Court decisions in a number of sensitive matters [100].

Another Cybercrime against persons is Cyberstalking. In actual life stalking and harassments are the problems that many persons especially women, are familiar with. The Oxford dictionary defines stalking as "pursuing stealthily" [101]. Cyber stalking is the process of following and observing a moments and activities and the behaviour of a person on the Internet by posting messages on the social media, bulletin boards by the victim, constantly bombarding the victim with Facebook messenger etc. In terms of internet, these problems are known as "Cyberstalking" or "on-line harassment" [102].

The second category of Cybercrimes is the Cybercrimes against property such as Financial Crimes. These crimes include unauthorized trespassing by digital systems through cyberspace, propagation of harmful programs and unauthorized ownership of computerized information. There are several kinds of financial crimes like hacking into bank servers, social engineering, cyber cheating, credit card frauds, accounting scams, cloud bursting etc. [103]. According to criminal law, fraud is the crime done intentionally by some adversary to deceive naive people in order to damage him/her and to obtain the benefit of their service unjustly. In the criminal law of common law jurisdictions, fraud is often referred to as theft by deception, larceny by trick, larceny by fraud and deception, or anything similar. There were two chronological steps to fraud. The first was deception by person A directed at person B, followed by a self-injurious action by person B. These steps were discussed in the 1950 case of Kat v Diment arising under the Merchandise Marks Act 1887 (UK). This statute was one in a series of criminal statutes, ultimately used and interpreted by Lord Diplock in the Advocaat Case, discussing the tort of passing-off, a tort of misrepresentation. Creation and propagation of destructive programming scripts which harm other computer systems is also Cybercrime. Violating the piracy of software programs is also a kind of Cybercrime. Any act by which the person is accessing the resources for which he/she is not authorized is an offence.

The third category of Cybercrime is the crime against Government and the society at large. One serious class of crime in this category is the Cyber Terrorism. Internet is being used by a single person or a group of people to make threats to the government organizations and also to terrorise the citizens [104]. The use of information technology by terrorist organizations is not limited to running websites and research in databases. After the investigations of 9/11 attacks, it was reported that the terrorists used e-mail to communicate and launch their attacks [105]. Cyber crime is manifested into cyber terrorism when an individual cracks an act of conscious, large-scale interruption of computer networks maintained by military or government organization. As per the section 66F of the Information Technology Act 2000 there is a tough punishment for cyber terrorists. The section 66F highlights unauthorized access to a computer resource with intention to threat the unity, integrity, security or sovereignty of India. As per this act there is punishment for cyber terrorism which may extend to "Life Imprisonment".

Cyber laws are usually framed by government's IT and cyber experts in every country. As per ministry of electronics and information technology, government of India, destroying or altering the computer network, computer program or computer source code is punishable with imprisonment up to three years, or with fine which may extend up to two lakh rupees, or with both [106]. Since there are very strict regulations and punishments for different category of cyber criminals but still this cyber crime especially cyber terrorism exists in every nation. Adversaries are easily hacking into banking systems, social networking websites, e-commerce websites, etc. [107]. Also India is reported as one of the top countries with the fake Facebook accounts[16]. In most of the cases, investigators were not able to get trace of the criminals as the crime was conducted across the boundaries of the nation. The tools and instruments needed to investigate cybercrime is quite different from those used to investigate ordinary crimes. Cyber law needs to be strengthened to handle cybercrimes across the boundaries otherwise we are likely to get unacceptable results. Security firewalls should be strong enough to stop intruders getting into the systems, IDSs (Introduction Detection Systems) should be strengthened enough to recognize the behaviour of attackers, OSN immune systems should be developed to indentify the anomalies and mitigate their effect on the overall network. However the Information Technology Act in India has proved to be inadequate to a certain extent during its application. The need of the hour is a worldwide uniform cyber law to combat cyber crimes. Cyber crime is a global phenomenon and therefore it should be tackled on the same level.

## 8. Simple remedies to mitigate fake profiles and Conclusions

A huge amount of users' personal, professional and even financial data is stored on several OSNs which attracts the attention of researchers, social analysts, data scientists etc. along with cyber criminals. Therefore, OSNs make the user's information vulnerable. Every OSN service provider is putting rigorous efforts to strengthen user security by imposing various constraints and measures for creating a new profile or making sure a user posts genuine contents on OSN. For example a user is supposed to create only one account per mobile connection or email address. Also at the same time a user is not allowed to clone any other person's account without any intimation to the original person [10], [76]. But there can be many accounts with the same name on a social network. The simple reason being various people with the same name may exist in real life then why not on social media. Furthermore, according to Facebook users using an account with somebody else's name, or maintaining a profile for any of his pet or maintaining a second profile that you use just for logging in to other sites, such an account is considered as fake. Parents are also violating rules by creating the profiles for their underage children because people under 13 are not allowed to have Facebook profiles. Several OSN service providers (for example Facebook, Twitter, Instagram, LinkedIn etc.) are taking users' privacy and security issue very seriously and are constantly seeking to enhance and strengthen their anti-counterfeit measures as mentioned in section 1.

The users of OSNs are provided with several options to report a spam or malicious contents or an abusive statement updates. For example on Facebook, the users are provided with options to report any content that is a spam or abusive or obscene [81]. Users have the privilege to report about any profile, post, message, page, group, event, or comment etc. [80]. Facebook has its own immune system [27] to detect undesirable profiles on the network by giving its user an option to report a profile as fake. Similarly, Twitter [82] and LinkedIn [83] also allow their users to report a recognised spam or a fugitive content. Recently Facebook introduced an Artificial Intelligence based system called Deep Text Tool

---

[16] http://www.gadgetsnow.com/tech-news/Facebook-may-have-over-100-million-fake-accounts-globally/articleshow/34672084.cms

[84] which is able to understand the text like humans. Besides helping the users with what they want to say, the tool would also be able to help in filtering the spam content in near future. Furthermore, Instagram is developing an anti-harassment tool [86] to filter comments or even help users to turn off the comment option on particular post. This way the users can block a particular comment to avoid harassment. Although OSN service providers are leaving no stone unturned to enhance the privacy and security of users, but there is still need to do something more than enabling users to just report a shady profile as fake or using the security and privacy features provided by different OSNs. With thousands of fake profile users on different OSNs having multifaceted aims to deceive, one need to adapt more advanced methods to secure one's online presence as least can be done when the security gets compromised.

In this paper we highlighted various types of OSN threat generators like compromised profiles, cloned profiles and online bots as fake profiles (spam-bots, social-bots, like-bots and influential-bots). A concise list of several categories of features used till date in literature to identify the fake profiles has been presented. The paper also highlights different data crawling approaches along with some existing data sources. A survey of techniques used by various researchers is presented for fake profile detection. A sincere effort has been taken to put everything about malicious entities existing on OSNs at one place. Many researchers have tried to mitigate the adverse effect of fake profiles to some extent but more concrete steps are still to be taken. Also a brief outlining of pro and cons of several existing cyber laws to curb the online fake profiles has been highlighted. It can be concluded that the need for more advanced automated methods still remains unfulfilled for secure social networks. The appropriate and timely steps are needed to develop automated mechanisms to identify suspicious users.